\documentclass[english,aps,prb, superscriptaddress]{revtex4-2}
\usepackage{babel}
\usepackage{amsmath}
\usepackage{amssymb}
\usepackage{graphicx}

\begin{document}
\title{Higher Chern Number States in Curved Periodic Nanowires}
\author{Zhuo Bin Siu}
\email{elesiuz@nus.edu.sg}
\affiliation{Department of Electrical and Computer Engineering, National University of Singapore, Singapore}
\author{Seng Ghee Tan} 
\affiliation{Department of Optoelectric Physics, Chinese Culture University, 55 Hwa-Kang Road, Yang-Ming-Shan, Taipei, Taiwan}
\author{Mansoor B.A. Jalil}
\email{elembaj@nus.edu.sg}
\affiliation{Department of Electrical and Computer Engineering, National University of Singapore, Singapore}

\begin{abstract}

The coupling between the spin and momentum degrees of freedom due to spin-orbit interactions (SOI) suggests that the strength of the latter can be modified by controlling the motion of the charge carriers. In this paper, we investigate how the effective SOI can be modulated by constraining the motion of charge carriers to curved waveguides thereby introducing real-space geometric curvature in their motion. The change in the SOI can in turn induce topological phase transitions in the system. Specifically, we study how the introduction of periodic sinusoidal curvature in nanowires with intrinsic SOC can induce the onset of mid-gap topologically protected edge states, which can be characterized by a topological invariant or Chern number. The Chern number corresponds to the number of discrete charges that would be pumped across the length of the nanowire when the phase of a sliding gate potential relative to that of the sinusoidal curvature is varied adiabatically over a complete period. In addition, coupling to an external magnetization can be utilized as an experimental knob to modify the Chern number by displacing the energies of the curvature-induced bands relative to one another. The magnetization can be tuned to achieve large discrete jumps in the number of pump charges per phase period.
\end{abstract}

\maketitle
\section{Introduction}

The dependence of the spin-orbit interactions (SOI) on the momentum of the charge carriers suggests that the SOI can be manipulated by introducing (real-space) geometrical curvature into the (constrained) motion of  charge carriers so that the direction and magnitude of their momentum vary with their positions \cite{Spin3_134006}. The modification of the effective SOI and appearance of curvature-induced terms \cite{PRB84_195457,PRB66_033107} due to the curved motion has led to various interesting phenomena, such the modification to the spin texture \cite{PRB87_174413,PRB94_081406}, spin precession \cite{PRB83_115305,PRB84_214423}, and charge density \cite{PRB84_085307}, suggesting potential applications for the generation of spin-polarized currents \cite{PRB75_085308, JAP115_17C513}. Recently, the independent geometrical control of the spin and charge resistances in a curved nanosystem has been demonstrated experimentally \cite{NanoLett19_6839}. In particular, a Brillouin zone can be defined and a miniband structure formed over the resulting superlattice when the curvature is periodic \cite{Nanotech27_135302, CondMat4_3, PLA383_2124, JPD49_295103, PRB79_235407}. With the introduction of SOI, the change in the direction of motion of the charge carriers as they move on a periodically curved waveguide lying on a flat surface can, for example, induce a finite  out-of-plane spin accumulation even when the SOI field is completely in-plane \cite{JAP121_233902}.  

The important role that SOI has historically played \cite{Science314_1757, Science318_766} in the study of topologically non-trivial states in condensed matter and the fact that topological invariants are quantities defined over a Brillouin zone \cite{PRL49_405, PRL95_146802} suggest that the introduction of a periodic curvature can induce topological phase transitions in an otherwise topologically trivial SOI system \cite{RMP82_3045, Nat464_194, RMP83_1057}. We have earlier shown that the introduction of a curvature into a four-band system can induce a topological phase transition accompanied by the emergence of edge states \cite{SciRep8_16497}. Ortix and co-authors have also investigated a one-dimensional sinusoidally curved nanowire with SOI, and showed that topological mid-gap edge states can emerge in such a system \cite{PRL115_256801}. In the presence of adiabatically rotating magnetizations, the non-trivial topological properties of the curved SOI nanowire can act as a charge pump \cite{PRB97_241103, PRB100_075402}. 

In this work, we consider a similar one-dimensional sinusoidally curved nanowire with SOI. To form a tunable superlattice structure, we add a periodic gate potential profile with the same periodicity as that of the nanowire geometry. We also introduce a magnetization coupling, which serves to modulate the sub-band structure and its topological properties. We show that the system harbours a topological invariant or Chern number that depends on the phase difference $\phi_{\mathrm{U}}$ between the sinusoidal curvature and the gate potential profile, which can be varied with time. This implies possible charge pumping across the nanowire by adiabatic variation of $\phi_{\mathrm{U}}$ with a discrete number of charges, i.e., the Chern number, being pumped per period. The Chern numbers \cite{JAP110_121301, PhyRep882_1} can be tuned by varying the magnetization coupling to realize bands with large Chern numbers. The magnetization can thus act an experimentally tunable knob for controlling the number of charges pumped per period.
\section{Methods}
We consider a one-dimensional periodically curved nanowire confined on the $xz$ plane with a SOI coupling strength of $\alpha$, an applied magnetization coupling $M_{\mathrm{y}}$ in the $y$ out-of-plane direction, and a periodic gate potential $U(x)$ described by the Hamiltonian
\begin{equation}
	H = \frac{1}{2m^*}\hat{p}^2 + \alpha \vec{\sigma}\cdot\hat{p} + M_{\mathrm{y}}\sigma_y + U(x) + U_{\mathrm{DC}} \label{Ham0} 
\end{equation}
where $\hat{p}$ is the momentum operator, the $\sigma$s are the spin operators, and $U_{DC}$ is the da Costa confinement potential \cite{PRA23_1982, AnnPhy63_586}, which exists intrinsically for a quantum mechanical particle confined to move on a curved surface. Several techniques, such as adhering the nanowire on prestrained substrates \cite{NatNano1_201} or electron beam lithography \cite{PRB82_155458} can be used to fabricate nanowires with the required curved geometries. 

Specifically, we assume a sinusoidal curvature with an amplitude of $z_0$ and a period of $\lambda$ so that the profile of the nanowire is given by 
\begin{equation}
	\vec{r} = x e_x + z_0 \sin( (2\pi/\lambda)x) )e_z \label{r}
\end{equation} 	
where $e_x$ and $e_z$ are the unit vectors along the Cartesian $x$ and $z$ directions. The nanowire is shown schematically in Fig. \ref{gFig1}a.  $\vec{r}$ defines the tangent vector to the nanowire $\vec{e}_{\mathrm{t}}$ via $\vec{e}_{\mathrm{t}} \equiv \partial_x \vec{r}$. The nanowire is shown schematically in Fig. 1a. 
The determinant of the metric tensor is conventionally denoted as $g$ and is given by 
\begin{equation}
	g = |\vec{e}_{\mathrm{t}}| = 1 + \left( \frac{ (2\pi z_0 \cos( (2\pi/\lambda)x))}{\lambda} \right)^2. \label{gDet}
\end{equation}
In one dimension, the da Costa confinement potential is given by  \cite{PRA23_1982}
\begin{equation}
U_{\mathrm{DC}} = -\frac{1}{2m^*}\left(\frac{(\partial^2_x\vec{r}\cdot\hat{n})}{2}\right)^2
\end{equation} 
where the normal vector $\vec{n} \equiv (\partial_x\vec{r})\times\hat{y}$.
For our nanowire, $ U_{\mathrm{DC}}$ has the explicit form of 
\begin{equation}
	U_{\mathrm{DC}} = -  \frac{ (2\pi^2 z_0\sin((2\pi/\lambda)x))^2}{2m^*(\lambda^2 + (2\pi z_0 \cos((2\pi/\lambda)x)^2))^3}.
\end{equation}

We assume that $U(x)$ has same periodicity as the curvature but with a phase offset of $\phi_{\mathrm{U}}$, i.e., 

\begin{equation}
	U(x) = U_0\sin( (2\pi/\lambda)x + \phi_{\mathrm{U}}).
\end{equation}
The Hamiltonian acting on an arbitrary wavefunction $\psi(x)$ is explicitly given by 
\begin{eqnarray}
	H\psi(x) &=& -\frac{1}{2m^*} \frac{1}{\sqrt{g}} \partial_x ( \frac{1}{\sqrt{g}}\partial_x \psi(x)) \nonumber \\
	&& - i\alpha(\vec{e}_t - \frac{1}{2}(\partial_x\vec{e}_t))\cdot\vec{\sigma}\partial_x \psi(x) \nonumber \\
	&& + (M_{\mathrm{y}}\sigma_y + U_{\mathrm{DC}} + U(x))\psi. \label{HamEx}
\end{eqnarray}
The first term on the right of the equal sign comprises the kinetic energy, the second term the spin-orbit interaction, and the third term the potential energy terms due to the applied potential, da Costa potential, and applied magnetization coupling. We obtain the eigenstates and eigenvectors of the Hamiltonian through direct numerical diagonalization of the real-space finite-difference approximation of the Hamiltonian Eq. \eqref{HamEx}, which would be described in detail in the following sub-sections. 

In curvilinear coordinates, the Hermiticity condition for an operator $O$ is expressed as   is $\langle \Psi|O|\Phi\rangle = \langle \Phi|O|\Psi\rangle^*$, where $|\Psi\rangle$ and $|\Phi\rangle$ are two arbitrary states, and
\begin{equation} 
	\langle \Psi|O|\Phi\rangle = \int \mathrm{d}x\ \sqrt{g} \Psi^*(x)O\Phi(x). \label{HermitCond} 
\end{equation} 
Note that there is an additional factor of $\sqrt{g}$ in the definition of the inner product due to the curvature in the spatial integration. To obtain the physical eigenstates of a lossless / gainless system, the Hamiltonian of the system has to be Hermitian. Thus, in the matrix representation of the Hamiltonian $\mathbf{H}$ (boldface here representing matrices), the $(i,j)$th  matrix element $\mathbf{H}_{ij}$ should satisfy $\mathbf{H}_{ij} = \mathbf{H}_{ji}^*$. Putting $O = H$ where $H$ is the Hamiltonian and $|\Psi\rangle = |\Phi\rangle = |x\rangle$ into Eq. \ref{HermitCond} suggests that instead of the usual Schr\"{o}edinger equation $\langle \vec{r}| H|\psi\rangle = \langle \vec{r}|\psi\rangle E$, we should consider a modified version given by $\langle x| \sqrt{g}H |\psi\rangle = \langle x|\psi\rangle \sqrt{g}E$. In other words, we diagonalize $\sqrt{g}H$ instead of $H$. The details of the construction of the finite difference approximation of the kinetic energy and SOI terms in Eq. \eqref{HamEx} are given in the subsections below. As for the remaining terms, the construction of the finite difference approximation is relatively straightforward.

\subsubsection{Kinetic energy operator} 
Let us first consider the generalized kinetic energy operator acting on arbitrary state $\psi$ at position $x$ :   
\begin{eqnarray}
	\sqrt{g}\psi^* \hat{p}^2\psi  &=& -\sqrt{g} \psi^* \frac{1}{\sqrt{g}} \partial_x ( \frac{1}{\sqrt{g}}\partial_x \psi)  \nonumber \\
	&=& -\psi^* \partial_x (\frac{1}{\sqrt{g}}\partial_x \psi) \nonumber \\ 
	&\stackrel{\mathrm{IbP}}{=}& (\partial_x \psi^*)\frac{1}{\sqrt{g}}(\partial_x \psi) \nonumber \\
	&=& \langle \psi| (\partial_x |x\rangle)\frac{1}{\sqrt{g}} (\partial_x \langle x|) |\psi\rangle \label{p2}  
\end{eqnarray}
where the result in the third line is obtained by integration by parts (``IbP''). 
 
With reference to the standard finite difference approximation $\partial_x |x\rangle \approx \frac{1}{2a} (|x+a\rangle - |x-a\rangle )$ where $a$ is the lattice spacing, we obtain the following approximation: 
\begin{align}
	& (\partial_x |x\rangle)( \partial_x \langle x|) \nonumber \\
  \approx& \frac{1}{2a^2} \Big ( ( |x+a\rangle - |x\rangle )(\langle x+a| - \langle x|) + (|x\rangle - |x-a\rangle)(\langle x| - \langle x-a|)  - (|x+a\rangle\langle x+a| + |x-a\rangle\langle x-a|) \Big) \nonumber \\
	=&\frac{1}{2a^2} ( -|x+a\rangle\langle x|-|x-a\rangle\langle x|-|x+a\rangle\langle x| - |x-a\rangle\langle x| +  2(|x\rangle  \langle x|) + |x+a\rangle\langle x+a| + |x-a\rangle\langle x-a| ). \label{dxpdxp} 
\end{align}
Using Eq. \eqref{dxpdxp}, Eq. \eqref{p2} reduces to the familiar form of
\[
	-|x\rangle \partial_i^2 \langle x| = -\frac{1}{a^2}|x\rangle ( \langle x+a| + \langle x-a| - 2\langle x|) 
\]
on a flat surface in which only immediately neighboring lattice sites are coupled together. 
Note that the values of $\frac{1}{\sqrt {g}}$ are generally different at lattice sites $x$ and $ x \pm$ a for a given curved surface. Therefore, in applying the approximation of Eq. \eqref{dxpdxp} in Eq. \eqref{p2}, we take the average of the values of $\frac{1}{\sqrt {g}}$ at the neighbouring bra and ket lattice sites, i.e.,  
\begin{align}
&\sqrt{g}\psi^*(x)\hat{p}^2\psi(x) \nonumber \\
 \approx& \frac{1}{2a^2}\Big[  - (\sqrt{g(x+a)}+\sqrt{g(x)}) \psi^*(x+a)\psi(x) \nonumber \\
& – (\sqrt{g(x-a)}+\sqrt{x})\psi^*(x-a)\psi(x) + 2\sqrt{g(x)}|\psi(x)|^2 \nonumber \\
& + \sqrt{g(x+a)}|\psi(x+a)|^2 + \sqrt{g(x-a)}|\psi(x-a)|^2 \Big].
\end{align} 
This procedure ensures the numerical Hermiticity of $\sqrt{g}H$, i.e., the numerical matrix representation of $\sqrt{g}H$ satisfies $\langle x\pm a|\sqrt{g}\hat{p}^2|x\rangle=\langle x|\sqrt{g}\hat{p}^2|x\pm a\rangle^*$.
\subsubsection{Spin orbit interaction} 
For a generic SOI interaction of the form $\vec{S}\cdot\vec{p}$ where $\vec{S}$ contains the spin operators and may in general be position-dependent, the symmetrization of the SOI Hamiltonian
\begin{equation}
	\sqrt{g} \frac{1}{2} (S^ip_i + p_iS^i) \simeq \frac{1}{2a} \sum_x  \left( \frac{1}{2}((\sqrt{g}S^i)_{[x]} + (\sqrt{g}S^i)_{[x + a]})   (|x \rangle \langle x + a| + |x + a\rangle \langle x| )\right) \label{genSOC} 
\end{equation} 
where $(\sqrt{g}S^i)_{[x]}$  ($(\sqrt{g}S^i)_{[x+a]}$) denotes the value of $\sqrt{g}S^i$ at lattice point $x$ ($x+a$) yields identical results to the standard da Costa dimensional reduction procedure \cite{PRA23_1982, SciRep8_16497} for a linear-momentum SOI operator on a curved surface. Here, we assume a SOI of the form of $\vec{p}\cdot\vec{\sigma}$, which has same form as the linear Dresselhaus SOI or the Rashba SOI\cite{JPCM30_285502} after a spin rotation. In this case, $S^ip_i\psi = (e_t\cdot\vec{\sigma})/g (-i\partial_x \psi)$, and
\begin{align}
	\sqrt{g}\frac{1}{2}(S^ip_i + p_iS^i) \simeq \frac{1}{{4a}} \sum_x \Big[ & \big( (\sqrt{g(x)}+\sqrt{g(x+a)})\sigma_x + (2\pi/\lambda)( \cos(2\pi/\lambda x)+ \cos(2\pi/\lambda (x+a)))\sigma_z\big) \nonumber \\
& \big(|x\rangle\langle x+a| + |x+a\rangle\langle x| \big) \Big]
\end{align} 
where $g(x)$ is the value of the metric tensor determinant Eq. \eqref{gDet} at the coordinate $x$ along the nanowire. 
\section{Results and Discussion} 
Unless otherwise stated, we set $m^*=0.02 m_e$ where $m_e$ is the free electron mass and $\alpha = 0.1\ \mathrm{eV nm}$, which are typical values for semiconductor Rashba two-dimensional electron gases \cite{PRL78_1335,PRB41_7685}. The wavelength of the curvature period is set at $\lambda = 200\ \mathrm{nm}$. 
\begin{figure}[htp]
\centering
\includegraphics[width=0.6\textwidth]{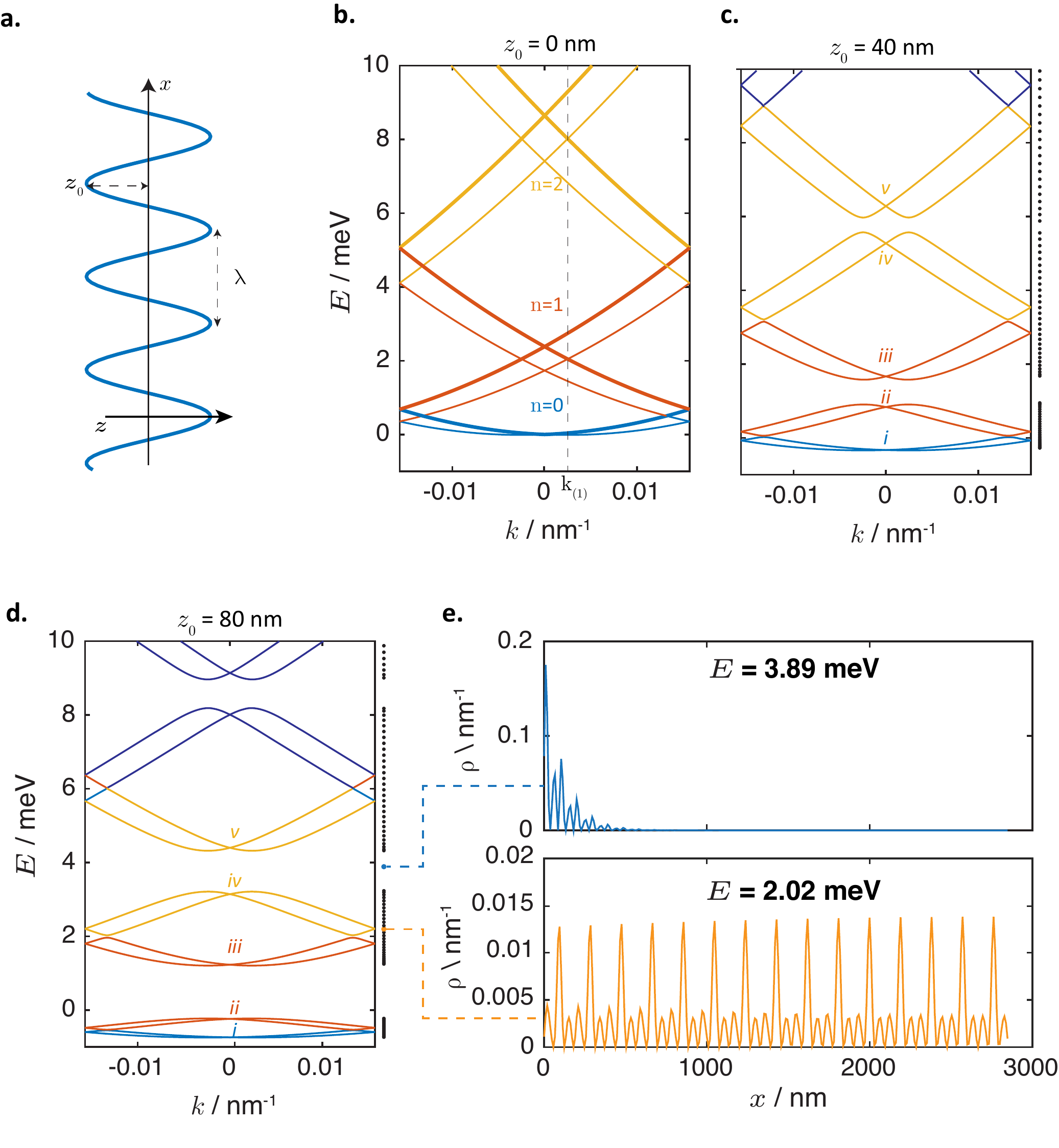}
\caption{  \textbf{a}. Schematic representation of sinusoidal nanowire lying on the $xz$ plane with the curvature amplitude of $z_0$ and curvature spatial period of $\lambda$. \textbf{b}. Band diagram of a flat nanowire in the reduced zone scheme where the period was taken to be $\lambda=20\ \mathrm{nm}$. The thick (thin) lines denote the states from the positive (negative) eigenspinor branch with energies $(k_n)^2/(2m^*) + \alpha k_n$ ($(k_n)^2/(2m^*)-\alpha k_n$)  where $k_n = k_x + (n\pi)/\lambda$. The value of $n$ for each of the bands plotted is indicated by the color of the band, and the value of $n$ each color corresponds to is indicated on the plot. The dotted line at $k = k_{(1)}$ indicate one value of $k$ at which the bands cross in the flat nanowire and anticrossing occurs when the periodic curvature is introduced . \textbf{c}. (left) Band diagram of an infinite periodically curved nanowire with a smaller curvature amplitude of $z_0 = 40\ \mathrm{nm}$. The Roman numbers i to v label the pairs of bands plotted in the same color as the corresponding Roman numbers. (right) The dots indicate the eigenenergies of a $15\lambda$ long finite nanowire with the same parameters. \textbf{d}. (left) Band diagram of an infinite periodically curved nanowire with a larger curvature amplitude of $z_0 = 80\ \mathrm{nm}$. The dots indicate the eigenenergies of a 15$\lambda$ long finite nanowire with the same parameters. \textbf{e}. The spatial densities of the $E=3.90\ \mathrm{meV}$ and $E=\ \mathrm{meV}$ eigenstates indicated by the blue and green dots in d.   }  
\label{gFig1}
\end{figure}	
\subsection{Eigenenergy spectrum of curved nanowire} 
We first investigate the effects of curvature on the eigenenergy spectrum of the nanowire in Fig. \ref{gFig1}. When the nanowire is completely flat (i.e., $z_0 = 0$) and the periodic potential is absent (i.e., $U_0=0$), the system becomes an isotropic SOI electron gas for which the eigenenergies are $\tilde{k}^2/(2m^*) \pm \alpha \tilde{k}$ where $\tilde{k}$ denotes that the wavevector in the extended zone scheme and can take values between $\pm\infty$. Although the flat nanowire is uniform, for the purpose of comparison, it can also be regarded as a periodic system with a period of $\lambda$ because $z(x) = z(x+\lambda) = 0$. This allows the introduction of a Brillouin zone with the boundaries located at $k = \pm\pi/\lambda$, where $k$ (without the tilde) denotes a wavevector in the reduced zone scheme. Note that the dispersion relations for $|\tilde{k}| > \pi/\lambda$ are folded back into the first Brillouin zone, as shown in Fig. \ref{gFig1}b. In Fig. \ref{gFig1}b,  the thicker dispersion curves correspond to the positive eigenspinor branch with energies of $k_n / (2m^*) + \alpha k_n$ where $k_n = k + (n \pi/\lambda)$. The fold-back of the dispersion curves to the reduced Brillouin zone leads to crossing points, such as $|k| = k_{(1)}$ between the $n$th positive and negative eigenspinor branch states, as indicated in the figure.

We now introduce a small curvature ($z_0 = 40\ \mathrm{nm}$) to the nanowire, and plot the resulting dispersion relation in Fig. \ref{gFig1}c. The introduction of a periodic curvature modulation to the nanowire opens large anticrossing band gaps at $|k|=k_{(1)}$, which mark the crossing points of the positive and negative eigenspinor branches in the absence of curvature. 
Moreover, the energies of the bands tend to be compressed and shifted downwards along the energy axis. For example, the two bands labelled as iii in Fig. \ref{gFig1}c, which span across an energy range of 1.6 meV, originate from the upper half of the $n = 1$ bands in Fig. \ref{gFig1}a. Originally, these span an energy range of 2.5 meV between 2 meV to 4.5 meV (Fig. \ref{gFig1}b), but now span a range of just 1.4 meV between 1.8 to 3.2 meV (Fig. \ref{gFig1}c). The compression becomes even more significant for the higher energy bands, as can be seen by comparing the positions of the band gaps in Fig. \ref{gFig1}c with the energies of the band crossing points at $k=k_{(1)}$ in Fig. \ref{gFig1}b.  

We now consider a periodic curved nanowire of a finite length along $x$. Because of the loss of translational invariance, $k$ is no longer a good quantum number, and the allowed eigenenergies of the finite-length wire become quantized owing to the quantum confinement effect. The series of dots depicted to the right of Fig. \ref{gFig1}c represent the discrete eignenergy spectrum of a nanowire of length $15\lambda$.  Note that the eigenenergies of the finite curved nanowire fall within the energies spanned by the energy bands of the infinite nanowire. In addition, there are no eigenstates of the finite nanowire that fall within the energy band gaps of the infinite  nanowire formed by the curvature-induced anticrossing. 

We will now demonstrate the onset of curvature-induced topological states in finite nanowires with periodic curvature of sufficiently large amplitude. Increasing the curvature amplitude $z_0$ results in a further compression of the infinite nanowire bands along the energy axis and a widening of the curvature induced anticrossing band gaps (Fig. \ref{gFig1}d). Interestingly, when the band gaps are sufficiently large, mid-gap states may emerge in the corresponding finite nanowire. The dots depicted to the right of Fig. \ref{gFig1}d correspond to the discrete eigenenergies of a finite nanowire of length $15\lambda$ at $z_0 = 80\ \mathrm{nm}$.. Note the presence of  (two) degenerate finite-length eigenstates at $E= 3.89\ \mathrm{meV}$ in the middle of the bandgap. Such mid-gap states are absent in the finite nanowire with a smaller curvature amplitude of 40 nm, as shown in Fig. \ref{gFig1}c. It is interesting to note that the mid-gap state is an edge state with localization near one end of the nanowire. This can be seen from the particle density plotted against the $x$ coordinate along the length of the nanowire in Fig. \ref{gFig1}e. For comparison, we plot the particle density for one of the finite-length eigenstates corresponding to $E=2.02\ \mathrm{meV}$  that lie within one of the bands (indicated by a green dot). In this case, the density profile shows a typical profile that extends throughout the interior of the nanowire. The presence of the mid-gap states with edge localization is an indication of the presence of non-trivial topological states in the curved nanowire, a point which we will subsequently prove by considering their corresponding Chern number.
\subsection{Effects of magnetization coupling and periodic potential on eigenenergy spectrum} 
\begin{figure}[htp]
\centering
\includegraphics[width=0.8\textwidth]{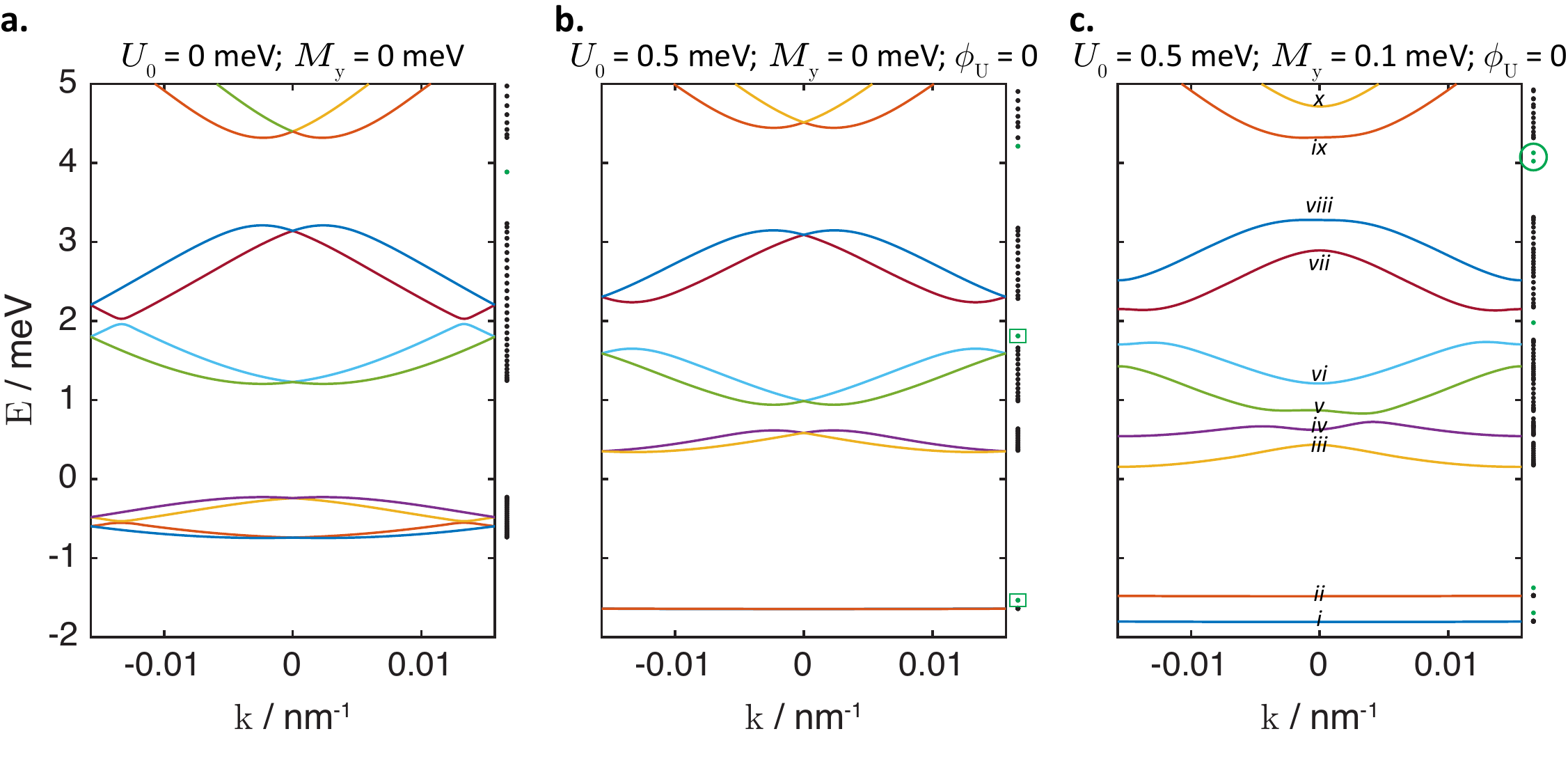}
\caption{  \textbf{a}--\textbf{c}. The dispersion relations of infinite periodically curved nanowires with $\lambda=200\ \mathrm{nm}$ and \textbf{a} $M_{\mathrm{y}}=0\ \mathrm{meV}$,  $U_0=0\ \mathrm{meV}$, \textbf{b} $M_{\mathrm{y}} = 0\ \mathrm{meV}$, $U_0 = 0.5\ \mathrm{meV}$, $\phi_{\mathrm{U}}=0$, and \textbf{c} $M_{\mathrm{y}} = 0.1\ \mathrm{meV}, U_0 = 0.5\ \mathrm{meV}$, $\phi_{\mathrm{U}}=0$, and the energy eigenvalues of the corresponding $15\lambda$ long finite nanowires with the same parameters. The colors of the bands denote the labels of the infinite-length nanowire from $i$ to $xi$ as indicated on in panel c. The eigenenergies of the finite nanowire mid-gap edge states are denoted by the green dots. The green squares in panel b indicate the new mid-gap edge states that emerge between panels a and b when the bandgaps are widened  by the introduction of the periodic sliding potential. The green circle in panel c indicates the pair of spin-split edge states between bands $viii$ and $ix$ in which the magnetization-induced energy splitting is the most evident after the introduction of the magnetization coupling.  }  
\label{gFig2}
\end{figure}

An adiabatic charge pump can be established by introducing a sinusoidal sliding potential $U_0$ and adiabatically varying the phase difference $\phi_{\mathrm{U}}$ between the sliding potential and the nanowire curvature. We will show that the amount of charge pumped per period as well as the number of edge states can be tuned by the magnetization $M_{\mathrm{y}}$. Fig. \ref{gFig2}a replicates the dispersion relation of the finite curved wire in Fig. \ref{gFig1}d so that the effects of the potential $U(x)$ and the magnetization can be more easily seen by comparison. Here, we have focused on a narrower energy range within which the edge states appear, and labelled the bands differently from Fig. \ref{gFig1}d to facilitate the discussion that follows. 

We will now impose a sliding potential on the nanowire with the same periodicity as its curvature, and show that this significantly modifies its bandstructure. The main plot of Fig. \ref{gFig2}b shows the bandstructure of an infinite nanowire in the presence of a sliding potential of $U_0 = 0.5\ \mathrm{meV}$ and zero phase difference ($\phi_{\mathrm{U}}=0$), while the discrete eigenstates of the corresponding finite nanowire of length $15\lambda$ are depicted to the right side of the plot. Comparing with the bandstructure in the absence of the sliding potential (Fig. \ref{gFig2}a), it can be seen that the sliding potential displaces pairs of bands along opposite directions along the energy axis – some pairs are pushed downwards, while others are pushed upwards. This causes a significant widening of the curvature-induced bandgaps. Generally, the sliding potential has a larger effect on lower-energy bands. For instance, the two upper-most bands (labelled $ix$ and $x$) are hardly affected by the presence of the potential, whereas there is a dramatic change and displacement of the four lowest bands (labelled $i$ to $iv$). This may be attributed to the relative size of the sliding potential compared with the energy of the bands. The sliding potential can also induce a band-gap opening, e.g., an energy gap is now opened between bands $ii$ and $iii$ in Fig. \ref{gFig2}b where previously no gap was present between the two bands shown in Fig. \ref{gFig2}a. Most interestingly, new mid-gap edge states (indicated by the squares in Fig. \ref{gFig2}b) also emerge in the energy gaps between bands $ii$ and $iii$, and between bands $vi$ and $vii$.

We now introduce magnetization coupling to the curved nanowire system as another means of modulating its bandstructure. Figure \ref{gFig2}c shows the bandstructure of the infinite nanowire and the eigenenergies of the finite-length nanowire in the presence of a magnetization coupling of $M_{\mathrm{y}} = 0.1\ \mathrm{meV}$ and the same sliding gate potential as was considered in Fig. \ref{gFig2}b. As before, the finite eigenstates of the corresponding finite nanowire with the same magnetization coupling and sliding potential are depicted as dots to the right of Fig. \ref{gFig2}c. The five pairs of bands in Figs. \ref{gFig2}a and \ref{gFig2}b originate from the SOI spin-split eigenspinor branches in the flat uniform nanowire system in Fig. \ref{gFig1}b, as we have explained earlier. For example, band $i$ ($ii$) originates from the negative (positive) eigenspinor branch of the $n=0$ bands of the flat nanowire. Note that for each pair, the bands are two-fold degenerate at $k=0$ where the SOI-induced splitting of the bands is zero. The application of the magnetization coupling lifts the degeneracy at $k=0$. For the finite curved nanowire, the energy degeneracy of the eigenstates is also lifted. Note that each of the dots in Fig. \ref{gFig2}a and \ref{gFig2}b actually represents two degenerate states and these are split in the presence of finite $M_{\mathrm{y}}$ coupling. Thus, there are now two distinct energy values of the mid-gap states between bands $ii$ and $iii$, $vi$ and $vii$, and $viii$ and $ix$ instead of a single energy value previously . (The energy splitting between the mid-gap states is most obviously seen in the $viii$ and $ix$ pair of mid-gap states in Fig. \ref{gFig2}c (circled) and is not readily visible at the scale of the figure for the remaining lower-energy pairs. ) Due to the energy splitting of the spin-split pairs arising from the magnetization coupling, alternate pairs of bands experience energy shifts of opposite signs. This opposite displacement in energy leads to a narrowing of the energy bandgaps, e.g., the gap between bands $iv$ and $v$. It is conceivable that as the magnetization coupling is increased further, the bands may cross each other in energy. Significantly, such band inversions are typically associated with topological phase transitions. We shall later show that this band inversion can indeed occur and result in a ``transfer'' of discrete Chern numbers between the bands that touch. Thus, magnetization coupling does not just act as an additional knob to modulate the energy dispersion of the curved nanowire system but can be applied to switch its topological properties.

\begin{figure}[htp]
\centering
\includegraphics[width=0.7\textwidth]{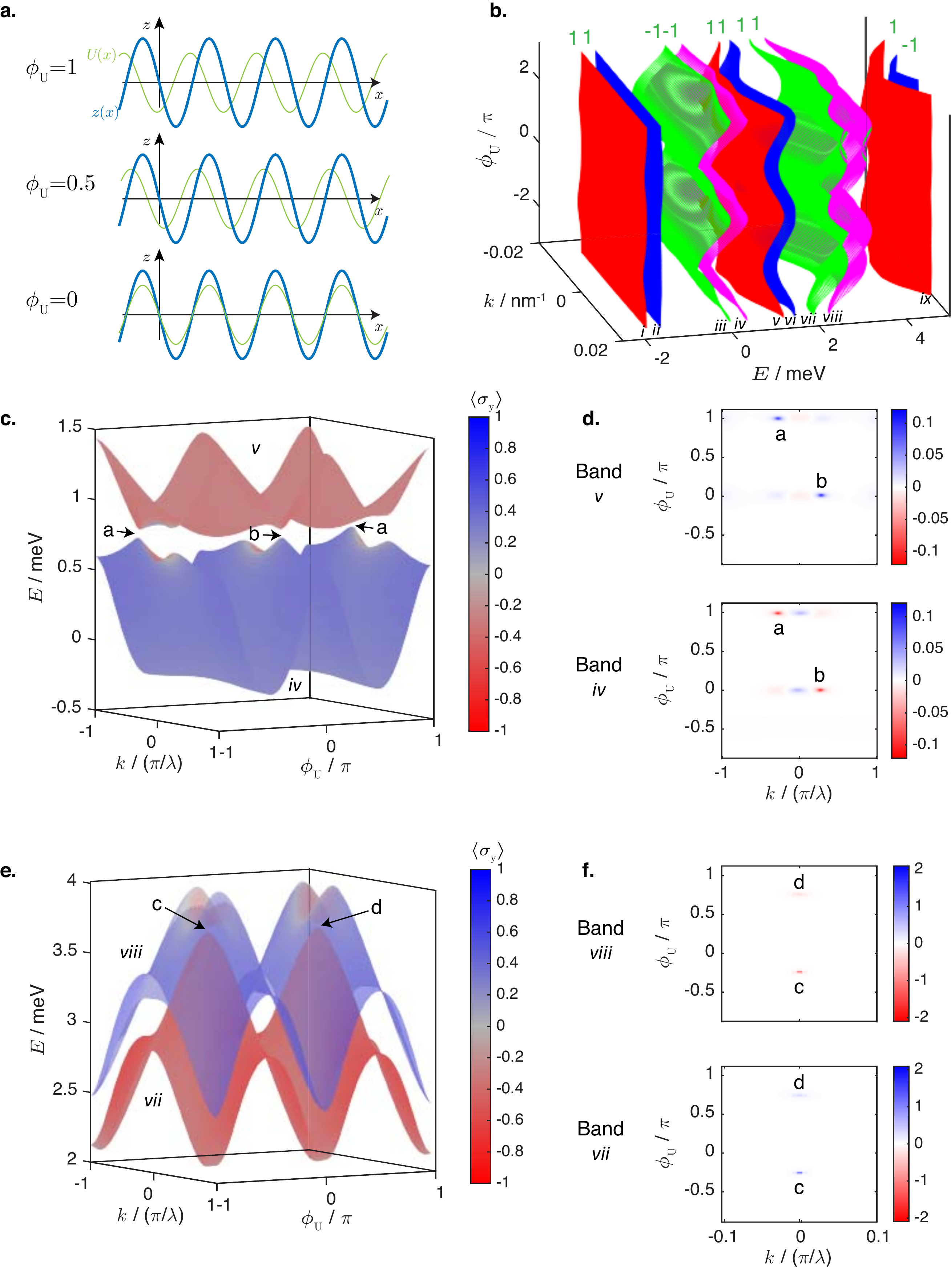}
\caption{  \textbf{a} Schematic diagram showing the relationship between the $xz$ profile of the nanowire (thick purple line) and the $x$-dependence of $U(x)$ at different values of $\phi_{\mathrm{U}}$. \textbf{b} The variation of the $E-k$ bandstructure relation of an infinite nanowire with $\phi_{\mathrm{U}}$ at $M_{\mathrm{y}}=0.1\ \mathrm{meV}$ and $U_0=0.5\ \mathrm{meV}$. Adjacent bands are plotted in different colors to make them easier to distinguish from one another. The numbers $\pm 1$ on the top of each band indicate the $k-\phi_{\mathrm{U}}$ Chern number of each band while the labels $i$ to $ix$ at the bottom of each band correspond to the labels $i$ to $ix$ in Fig. \ref{gFig2}. \textbf{c} The energies of bands $v$ and $vi$. The color of the eigenenergy surface at each value of $(k,\phi_{\mathrm{U}})$ for each band indicate the spin $y$ expectation value of the corresponding eigenstate. $a$ and $b$ denote the nearly-touching points between bands $v$ and $vi$. (The label for $a$ appears on both the left and right edges of the plots because $a$ is located on the edge of the  periodic Brillouin zone.)  \textbf{d} The distribution of the Berry curvature on the $k-\phi_{\mathrm{U}}$ plane in bands $v$ and $vi$. Note that the boundaries of the plots are slightly displaced from $\phi_{\mathrm{U}} = \pm \pi$ so that the nearly-touching point $a$ does not appear on the boundary of the plotted area. \textbf{e, f} The corresponding plots of the energies and Berry curvature distributions of bands $vii$ and $viii$. }
\label{gFF3} 
\end{figure} 
\subsection{Charge pumping and Chern number in curved nanowire system}	
Finally, we consider another parameter $\phi_{\mathrm{U}}$ which is the phase between the nanowire curvature $z(x)$ and the sliding potential $U(x)$. This allows us to define a two-dimensional parameter space in conjunction with the wavevector $k$, and thus characterize the topological properties of the curved nanowire system. Fig. \ref{gFF3}a schematically shows the displacement of $U(x)$ with respect to $z(x)$ as the phase difference as $\phi_{\mathrm{U}}$ is increased. Fig. \ref{gFF3}b shows how the bandstructures of the infinite nanowire system in Fig. \ref{gFig2}c vary with $\phi_{\mathrm{U}}$. Although the energy ranges spanned by the infinite wire bands may overlap as $\phi_{\mathrm{U}}$ is varied, the bands do not, in general, touch one another. The absence of energy degeneracy between the different bands in the $k-\phi_{\mathrm{U}}$ space allows us to define a topological Chern number in that space for each band as  
\begin{equation}
	n_b = \frac{1}{2\pi}\int^\pi_{-\pi} \mathrm{d}k \int^\pi_{-\pi} \mathrm{d}\phi_{\mathrm{U}}  \Omega_b(k,\phi_{\mathrm{U}})
\end{equation}
where $n_b$ is the $b$th band, and $\Omega_b(k,\phi_{\mathrm{U}})$ is the Berry curvature of this band, and is turn given by 
\begin{equation}
\Omega_b(k,\phi_{\mathrm{U}}) = 2\mathrm{Im}(\partial_k\langle \psi_b|)(\partial_{\phi_{\mathrm{U}}}|\psi_b \rangle).
\end{equation} 
Here, $|\psi_b\rangle$ is the eigenstate of the $b$th band and is a function of both $k$ and $\phi_{\mathrm{U}}$, and $n_b$ is its Chern number.  The resulting Chern numbers for each of the bands in Fig. \ref{gFF3}b are labelled on top of the band. 

It is necessary to understand the distribution of the Berry curvatures amongst the bands in order to  explain the variation of the band Chern numbers with the applied magnetization. We therefore focus on two pairs of bands, namely, bands $iv$ and $v$, and bands $vii$ and $viii$. The energies of these two pairs of bands are plotted in the surface plots in Fig. \ref{gFF3}c and Fig. \ref{gFF3}e, respectively, while the Berry curvature distributions of the corresponding bands on the $k-\phi_{\mathrm{U}}$ plane are plotted in Fig. \ref{gFF3}d and Fig. \ref{gFF3}f. Consider bands $iv$ and $v$ in Fig. \ref{gFF3}c first. There are two points labelled as $a$ and $b$ at which the two bands nearly touch  (the point $a$ appears at both the left and right edges of the plot because the band-touching point is on the Brillouin zone boundary in the plot). These points originate from the anti-crossing points labelled as $k_{(1)}$ in Fig. \ref{gFig1}b at which the successive bands are closest in energy from one another. Fig. \ref{gFF3}d shows that the Berry curvatures in bands $iv$ and $v$ are concentrated around these nearly-touching points, and that the Berry curvatures at a given point, say $a$, have opposite signs on the two bands. These trends can be explained by an alternative expression for the Berry curvature 
\begin{equation}
	\Omega_b =i \sum_{b'\neq b} \frac{  \langle b| (\partial_k H) | b'\rangle\langle b'|(\partial_{\phi_{\mathrm{U}}}H)|b\rangle - \langle b| (\partial_{\phi_{\mathrm{U}}} H) | b'\rangle\langle b'|(\partial_k H)|b\rangle }{(E_b – E_{b'})^2}. \label{altBC}
\end{equation}
It is evident that the amplitude of Berry curvature increases sharply as the energy difference between the bands decreases, and that the contributions of the band $i$ to the Berry curvature of band $i+1$ (i.e., the term being summmed in Eq. with $b=i$, $b'=i+1$) has an opposite sign from the contribution of band $i+1$ to the Berry curvature of band $i$. We can also observe in Fig. \ref{gFF3}f the same tendencies for the Berry curvatures to be concentrated at the nearly touching points ($c$ and $d$) between two adjacent bands $vii$ and $viii$  and the opposite signs of the Berry curvatures of the two bands. 
\begin{figure}[htp]
\centering
\includegraphics[width=0.8\textwidth]{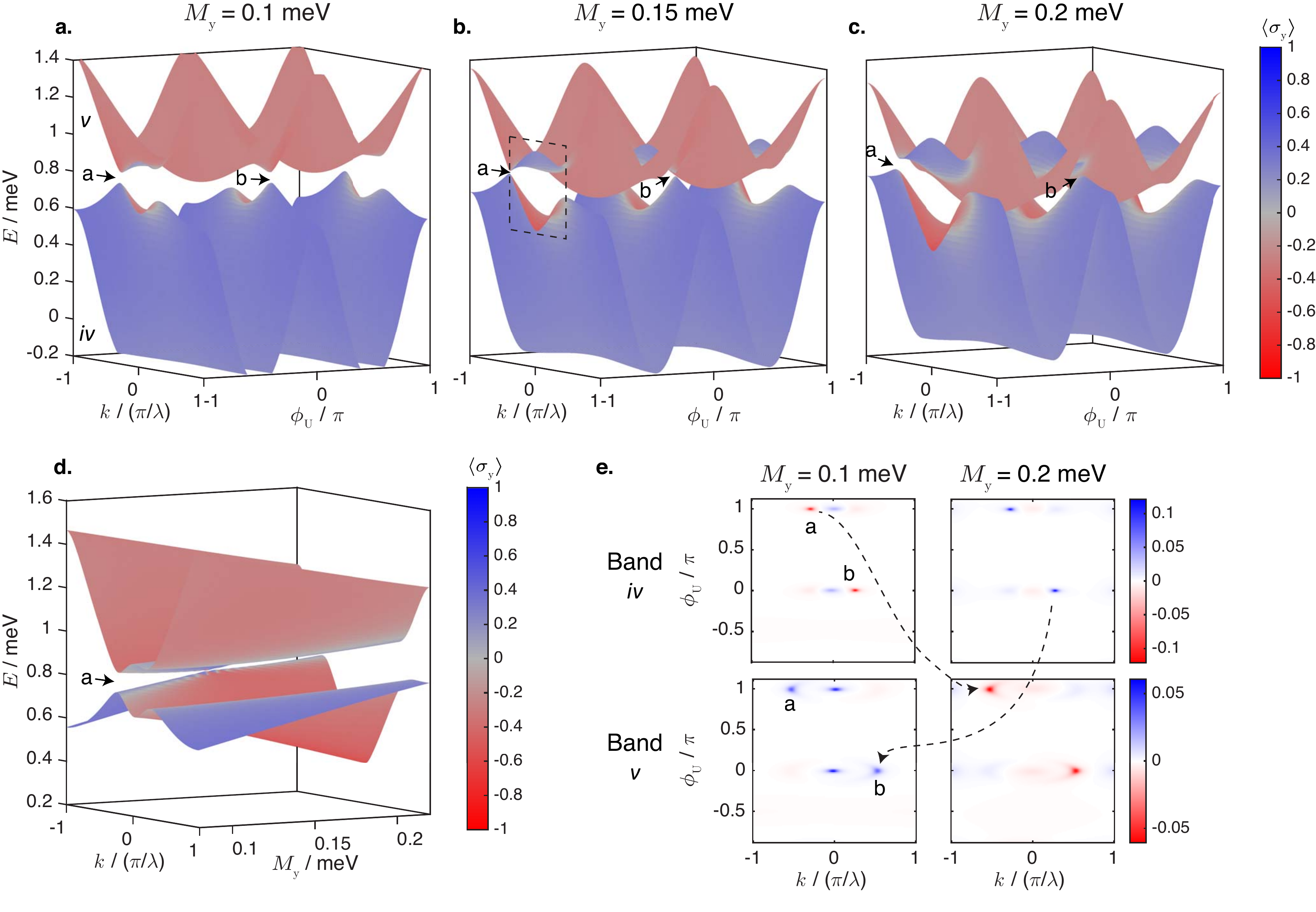}
\caption{  \textbf{a, b, c} The energies and spin $y$ expectation values of bands $iv$ and $v$ at a. $M_{\mathrm{y}} = 0.1$ meV, b. $M_{\mathrm{y}} = 0.15$ meV, and c.  $M_{\mathrm{y}} = 0.2$ meV. The dotted box in b. indicates a region on the energy surfaces where the two bands interpenetrate each other, as indicated by the reversal of sign of the spin $y$ expectation relative to the rest of the bands. \textbf{d} The energy and spin $y$ expectation values of bands $v$ and $vi$ at the $k_x$ value of point $a$ as functions of $M_{\mathrm{y}}$ and $\phi_{\mathrm{U}}$. The two bands touch at only a single isolated value of $M_{\mathrm{y}}$. \textbf{e}  The Berry curvature distributions of bands $iv$ and $v$ on the $(k, \phi_{\mathrm{U}})$ band at $M_{\mathrm{y}} = 0.1$ meV before the two bands touch at points $a$ and $b$, and at $M_{\mathrm{y}} = 0.2$ meV after the two bands have touched. The dotted arrow depicts the transfer of the negative Berry curvature at point $a$ from band $iv$ to $v$, and the transfer of the positive Berry curvature from band $iv$ to band $v$ at point $b$ after the bands have touched. } 
\label{gFF4} 
\end{figure} 

The distributions of the Berry curvatures across the bands change with the variation of $M_{\mathrm{y}}$ as adjacent bands touch and overlap with each other. We have seen earlier in Fig. \ref{gFig2}c that as $M_{\mathrm{y}}$ is increased, the bands polarized in the $+y$ direction are pushed up in energy, while those polarized in the $-y$ direction are pushed down in energy. This may then result in bands with opposite signs of spin polarization coming into contact with one another. Fig. \ref{gFF4} explains the manner in which the energies and Berry curvature distribution of two adjacent bands change when the bands come into contact. Fig. \ref{gFF4}a  reproduces the energy distribution of bands $iv$ and $v$ at $M_{\mathrm{y}} = 0.1$ meV shown in Fig. \ref{gFF3}c. The colors of the eigenenergy surfaces indicates the spin polarization of the bands. When $M_{\mathrm{y}}$ is increased to 0.15 meV, as shown in Fig. \ref{gFF4}b, band $iv$ (which has a largely positive spin $y$ polarization) is pushed upwards in energy, while band $v$ (which  has largely a negative spin $y$ polarization) is pushed downwards in energy, causing the two bands to approach each other.   The shifts in the energies of bands $iv$ and $v$ is perhaps most evident from the band bottom of band $iv$ and the band top of band $v$, which are cropped off in Fig. \ref{gFF4}a (notice that the bands terminate on the $k-\phi_{\mathrm{U}}$ planes at the top and bottom of the surface plots in straight lines) but are now visible in Fig. \ref{gFF4}b for the same energy range (notice that the top of band $v$ and bottom of band $iv$ now terminate in smooth curves in the range shown). Notably, there are also small regions centered around $k=0$ in the figure (one of which is indicated by the dotted box in Fig. \ref{gFF4}b), at which the spin $y$ polarization of the two bands differ from that of their surrounding regions.These regions can be interpreted as the interpenetration of the peak regions of band $iv$ into band $v$, and the trough regions of band $v$ into band $iv$ separated by anti-crossing energy gaps. 

These gaps may close at the nearly-touching points $a$ and $b$ with the variation of $M_{\mathrm{y}}$. A comparison between Fig. \ref{gFF4}a and Fig. \ref{gFF4}b shows that the energy gaps between bands $iv$ and $v$ at points $a$ and $b$ decrease with the increase of $M_{\mathrm{y}}$ from 0.1 meV to 0.15 meV such that and the two bands seem almost touch each other at points $a$ and $b$ in Fig. \ref{gFF4}b. A further increase in $M_{\mathrm{y}}$, as shown in Fig. \ref{gFF4}c for $M_{\mathrm{y}} = 0.2 $ meV, causes the two bands to further interpenetrate into each other,  as can be seen from the increase in the negative (positive) spin $y$-polarized region in band $iv$ ($v$). The energy separations between the two bands at points $a$ and $b$ have also increased compared to the separation in Fig. \ref{gFF4}b, indicating that the bands touch only at isolated values of $M_{\mathrm{y}}$. This is shown in Fig. \ref{gFF4}d,  where the $k$ is fixed to correspond to that of point $a$ while $M_{\mathrm{y}}$ is varied. It can be clearly seen that as $M_{\mathrm{y}}$ increases, the energy separation between bands $iv$ and $v$ initially decreases, becomes 0 (where the two bands touch at $M_{\mathrm {y}} = 0.1394 $ meV) , and then increases again.   An exchange of the Berry curvatures occurs between the bands in which the Chern number of one band increases by 1 and that of the other band decreases by 1 when the two bands  touch each other \cite{PRB79_195321}, as shown in Fig. \ref{gFF4}e. At $M_{\mathrm{y}}=0.1$ meV (left columns), the Berry curvatures around points $a$ and $b$ are negative in band $iv$, and positive in band $v$. The two bands touch at point $a$ at $M_{\mathrm{y}}=0.1394$ meV and at point $b$ at $M_{\mathrm{y}}=0.1414$ meV and swap their Berry curvatures at these touching points. Thus, at $M_{\mathrm{y}}=0.2$ meV, the Berry curvatures around points $a$ and $b$ become positive in band $iv$, and negative in band $v$ ( as shown in Fig. \ref{gFF4}e ), and the Chern number of the band $iv$ ($v$) correspondingly changes from -1 (+1) at $M_{\mathrm{y}}=0.1$ meV to +1 (-1) at $M_{\mathrm{y}}=0.2$ meV. 

\subsubsection{Energy dispersions of finite nanowires} 
\begin{figure}[htp]
\centering
\includegraphics[width=0.8\textwidth]{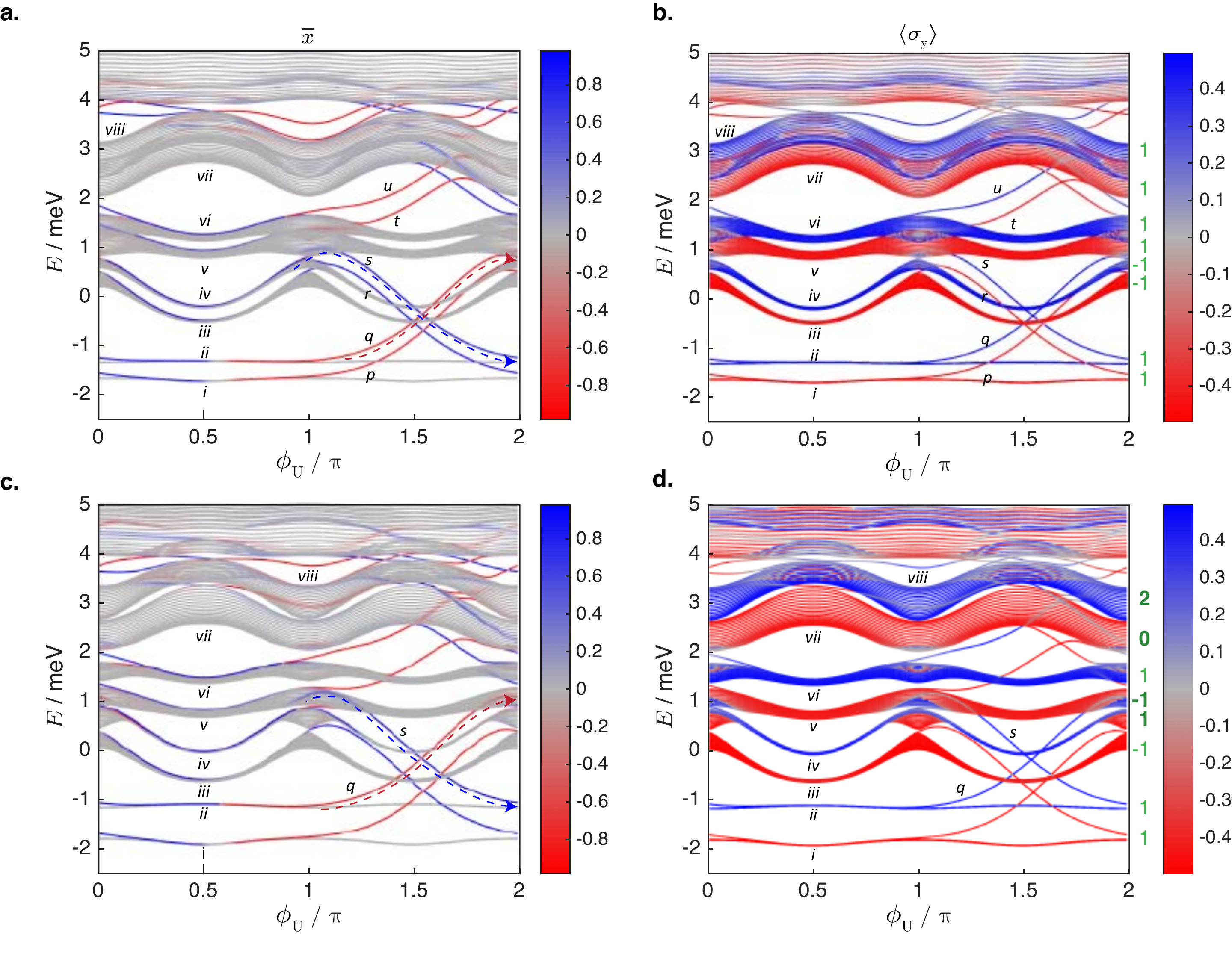}
\caption{ \textbf{a} The eigenenergies of a 20$\lambda$ long finite nanowire at $U_0 = 0.5\ \mathrm{meV}$ and $M_{\mathrm{y}} = 0.1\ \mathrm{meV}$ as functions of $\phi_{\mathrm{U}}$ and normalized $x$ expectation values of each eigenstate where a value of 1 (-1) indicates that the state is totally localized at the right (left) edge, and values near 0 indicate that the state is a bulk state. The labels from $i$ to $viii$ label the infinite-length nanowire bands each group of finite-length nanowire bulk states originate from. The labels $p$ to $u$ label the edge states. The dotted lines serve as guides for the eye to trace the evolution of the edge states with $\phi_{\mathrm{U}}$. \textbf{b} The eigenenergies and the spin $y$ expectation values of the eigenstates of the nanowire in a. The numbers at the right of the plot indicate the Chern numbers of the infinite wire bands from which the states originate. \textbf{c,d} The corresponding plots to a and b at $M_{\mathrm{y}}=0.2\ \mathrm{meV}$ } 
\label{gFF5} 
\end{figure}

The changes in the Chern numbers of the bands  as they come into contact with one another with the variation of $M_{\mathrm{y}}$ are reflected in the dispersion of the energies with $\phi_{\mathrm{U}}$ and the edge states that emerge in finite nanowires. In Fig. \ref{gFF5}a--d the eigenenergy dispersions are plotted as functions of $\phi_{\mathrm{U}}$ for finite nanowire systems of 20 periodic units. The colors of the lines Fig. \ref{gFF5}a indicate the normalized position $\bar{x} \equiv \langle \psi| [\hat{x} - (l/2)] |\psi\rangle / (l / 2)$ where $l$ is the length of the nanowire, $\hat{x}$ is the position operator, and $|\psi\rangle$ is an eigenstate of the finite-length nanowire which spans from $x=0$ to $x=l$. An eigenstate with a $\bar{x}$ value of nearly 1 (-1) therefore indicates a topological edge state which is localized near the right (left) edge of the nanowire  (e.g., the upper right plot in Fig. \ref{gFig1}d). Conversely, a $\bar{x}$ value of nearly 0 indicates a bulk state (e.g., the lower right plot in Fig. \ref{gFig1}d) where the probability distribution of the state is more symmetrically distributed about the centre of the nanowire. Plotting the energy dispersion of the finite nanowire with $\phi_{\mathrm{U}}$  also allows the origin of its energy bands to be traced more easily. For example, there are two distinct but overlapping groups of bulk states in the finite nanowire lying between 2 to 4 meV (Fig. \ref{gFF5}a). (We shall use ``band'' to refer to each band in the \textit{infinite} nanowire, and ``group'' to refer to the group of bulk bands in the finite wire that result from the quantization of each infinite wire band.)  These groups of bulk states can be traced to the bands $vii$ and $viii$ of the infinite nanowire (Fig. \ref{gFF4}b). Although the energy range spanned by the two bands of the infinite nanowire overlap, they do not actually touch anywhere in the Brillouin zone. Likewise, the corresponding bands in the finite nanowire of Fig. \ref{gFF5}a also overlap in energy but do not form any anticrossing. Consequently, we can label each group of bulk states of the finite nanowire in Fig. \ref{gFF5}a with the same Roman numeral index of the infinite nanowire band which they originate from, along with the Chern number of the band in Fig. \ref{gFF5}b.  

In order to further differentiate the different bulk bands of the finite nanowire as well as analyze the effect of the magnetization coupling $M_{\mathrm{y}}$, we plot the spin $y$ expectation value $\langle \psi|\sigma_y|\psi\rangle$  corresponding to each state (Fig. \ref{gFF5}b).  As we have described earlier, in the infinite nanowire, the effect of the $M_{\mathrm{y}}$ coupling is to further split the spin-split SOC bands, such that states with negative (positive) spin $y$ polarization will be pushed downwards (upwards) in energy. Hence, in the finite nanowire dispersion of Fig. \ref{gFF5}b, group $vii$ which has a negative spin $y$ polarization (reddish colour) is pushed below in energy compared to group $viii$ which has a positive $y$ polarization (bluish colour) relative to their original energies when the $M_{\mathrm{y}}$ coupling was absent. Thus, generally, the effect of $M_{\mathrm{y}}$ coupling produces pairs of groups of bands in the finite-length nanowire with alternating signs of the spin $y$ polarization, with the groups labelled by odd (even) Roman numerals being polarized along the negative (positive) spin $y$ polarization. 

The combination of finite Chern numbers and the presence of boundaries leads to the emergence of edge states in finite nanowires. These edge states are localized near the left (red lines) or right (blue lines) edge of the nanowire and are labelled $p$ to $u$ (Fig. \ref{gFF5}a). The energies of these edge states span the energy gaps between the groups of bulk states as $\phi_{\mathrm{U}}$ is varied from 0 to $2\pi$. For example, consider the edge states $q$ and $s$, which connect groups $ii$ and $iv$, in Fig. \ref{gFF5}a.  Within some range of values of $\phi_{\mathrm{U}}$, this edge states may penetrate into the bands $iii$ and $v$ of bulk states, respectively, but emerge from these bands more or less intact with the further increase of $\phi_{\mathrm{U}}$, before eventually merging into their destination bands. As $\phi_{\mathrm{U}}$  increases from 0 to $2\pi$, the edge state $s$, which is localized on the right edge, gets detached from band $iv$ and decreases in energy and eventually joining band $ii$.  Conversely, the edge state $q$, which is localized on the left edge, increases in energy and leaves band $ii$ to join band $iv$. The physical interpretation of these band-gap crossings by the pair of edge states is as follows: As $\phi_{\mathrm{U}}$ is adiabatically varied from $0$ to $2\pi$, a charge in edge state $ s$ localized on the right is removed from band $iv$ while a charge in edge state $q$ which is localized on the left is brought from band $ii$ to band $iv$ to replace it. This results in the net movement of a charge from the right edge of band $iv$ to the left edge as  $\phi_{\mathrm{U}}=2\pi$ is varied over a period (from 0 to $2\pi$).  Hence, band $iv$ thus has a Chern number of -1 (Fig. \ref{gFF5}b). Likewise, the Chern numbers of the other bands can be related to the transport of the edge states via the evolution of $\phi_{\mathrm{U}}$ over a complete period.

We now turn our attention to the effects of increasing the $M_{\mathrm{y}}$ coupling on the dispersion spectra as well as the topological properties of the bands. Figs. \ref{gFF5}c and Fig. \ref{gFF5}d are analogous to Fig. \ref{gFF5}a and  \ref{gFF5}b except that the magnetization coupling is increased to $M_{\mathrm{y}}$ = 0.2 meV. As we have seen earlier in Fig. \ref{gFF4}a and Fig. \ref{gFF4}c,  increasing $M_{\mathrm{y}}$ from 0.1 meV to 0.2 meV results in interpenetration between bands $iv$ and bands $v$. The interpenetration of the two bands is manifested by the sign reversal of the spin $y$ expectation values of the two corresponding groups of bands in the finite wire (comparing Fig. \ref{gFF5}b and Fig. \ref{gFF5}d) at $\phi_{\mathrm{U}}=0$  and $\phi_{\mathrm{U}}=\pi$, which respectively correspond to the projections of point $b$ and $a$ onto the $\phi_{\mathrm{U}}$ axis, respectively.  As we have also seen in Fig. \ref{gFF4}e, the increase in $M_{\mathrm{y}}$ changes the Chern numbers of groups $iv$ and $v$ in the finite wire. This change is reflected in the trajectories of the edge states among the bulk bands in the dispersion of the finite wire. For example, edge state $s$, which passes from group $iv$ and enters and subsequently exists group $v$ as shown in Fig. \ref{gFF5}a. At the larger value of $M_{\mathrm{y}} = 0.2$ meV, state $s$ is now completely contained within group $v$ (see Fig. \ref{gFF5}c) for $\phi_{\mathrm{U}} < \pi$ due to the exchange of the Chern numbers  between bands $iv$ and $v$ at the touching point $a$ on the $k-\phi_{\mathrm{U}}$ (see Fig. \ref{gFF4}e) . Similarly, edge state $q$, which terminates in group $iv$ at $\phi_{\mathrm{U}} = 2\pi$ in Fig. \ref{gFF5}a at the lower $M_{\mathrm{y}}$ value, now terminates inside group $v$ at  $\phi_{\mathrm{U}} = 2\pi$ at the higher $M_{\mathrm{y}}$ value (Fig. \ref{gFF5}c) due to the exchange of the Chern numbers between bands $iv$ and $v$ at point $b$ on the $k-\phi_{\mathrm{U}}$ plane (see Fig. \ref{gFF4}e).  

The exchange of the Chern numbers between adjacent bands also has a significant impact in the spin $y$ polarization of the eigenstates in Fig. \ref{gFF5}b and Fig. \ref{gFF5}d. At the lower $M_{\mathrm{y}}$ value of 0.1 meV, the edge state $s$ and $p$ emerge from the vicinity of the peaks of group $iv$ at $\phi_{\mathrm{U}}=\pi$ and $\phi_{\mathrm{U}}=2\pi$, respectively. The peak region of group $iv$ have a positive spin $y$ polarization. At the higher $M_{\mathrm{y}}$ value of 0.2 meV, the peak regions of group $iv$ have ``broken away'' from band $iv$ from band $iv$ and become embedded into group $v$ (denoted by blue regions in Fig. \ref{gFF5}d). The regions that have broken off carry along with them the end points of the state bands $s$ and $q$ along with them so that these states now terminate in group $v$ rather than group $iv$. Similarly, groups $vii$ and $viii$ have also undergone a change in Chern numbers and their related edge states when $M_{\mathrm{y}}$ was increased from 0.1 meV to 0.2 meV (compare Figs. \ref{gFF5}a and \ref{gFF5}b with Figs. \ref{gFF5}c and \ref{gFF5}d). Here, the upward (downward) shift of group $viii$  ($vii$) results to the emergence of a distinct band gap between the groups in which edge states now emerge. 

\subsubsection{Tuning of Chern numbers by magnetization magnitude and direction} 

\begin{figure}[htp]
\centering
\includegraphics[width=0.8\textwidth]{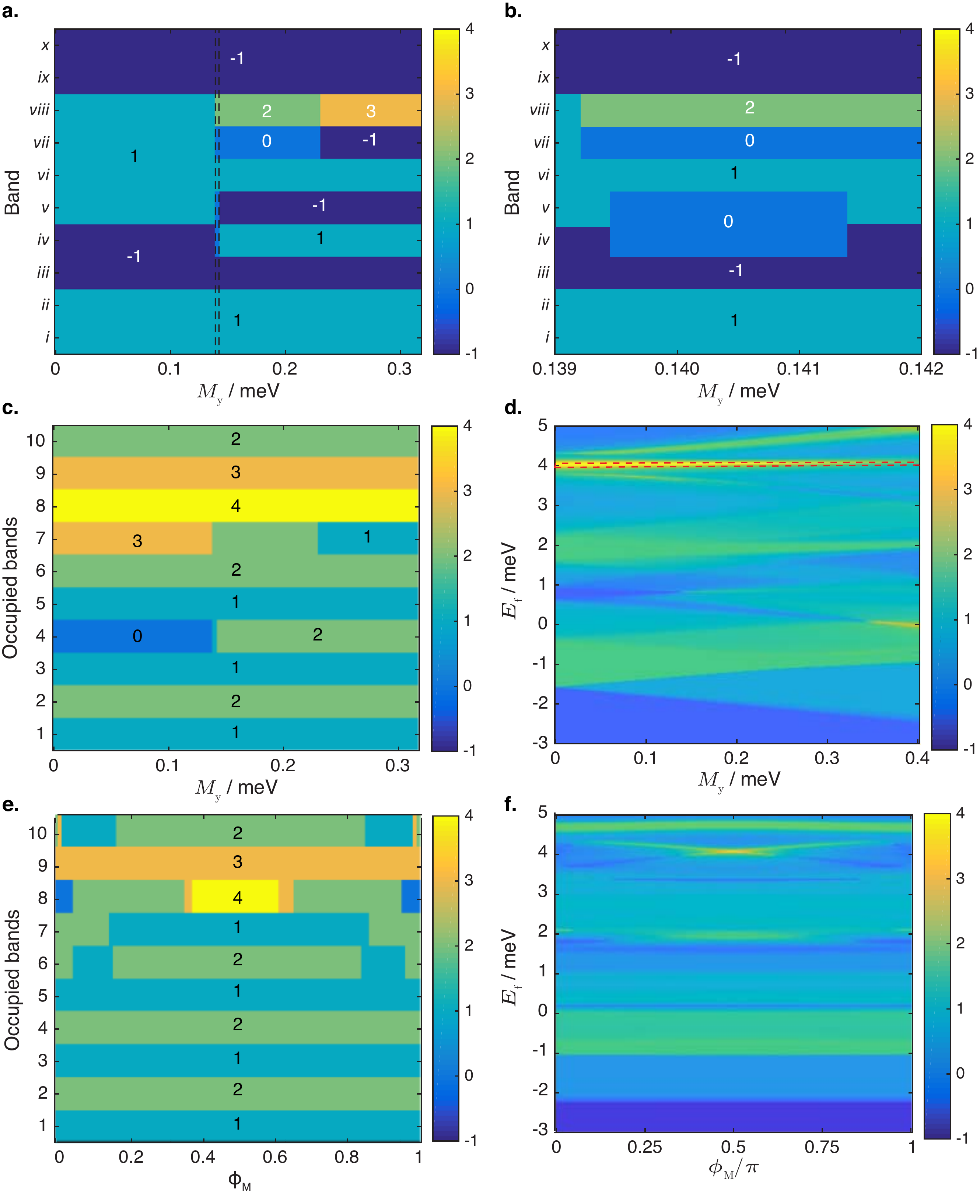}
\caption{  \textbf{a} Chern numbers of infinite-length nanowire bands arranged in order of increasing energy as functions of $M_{\mathrm{y}}$ for a nanowire system with the same parameters as Fig. \ref{gFF5}. \textbf{b}. An enlarged view of panel (a) for $M_{\mathrm{y}}$ between  0.139 meV to 0.142 meV as denoted by the dotted box.  \textbf{c} Cumulative Chern numbers of the same infinite nanowire summed over the occupied bands as functions of the number of occupied bands and $M_{\mathrm{y}}$. \textbf{d} Cumulative Chern numbers of the same system as b summed over the occupied bands as functions of the Fermi energy $E_{\mathrm{f}}$ and $M_{\mathrm{y}}$. The red dotted region corresponds to the region in the $E_{\mathrm{F}}-M_{\mathrm{y}}$ plane in which there are eight occupied bands. \textbf{e} Cumulative Chern numbers summed over the occupied bands as functions of the number of occupied bands in the same infinite-length nanowire except that the direction of the magnetization coupling is varied, i.e. $\hat{m} =  (\cos(\phi_{\mathrm{M}})\sigma_x + \sin(\phi_{\mathrm{M}})\sigma_y)$ and the magnitude of the coupling is fixed to 0.3 meV. \textbf{f} Cumulative Chern numbers of the same infinite nanowire in e summed over the occupied bands as a functions of the Fermi energy.} 
\label{gFig4} 
\end{figure} 

We further investigate the evolution of the Chern numbers with the variation in the magnetization $M_{\mathrm{y}}$ (see Fig. \ref{gFig4}a). The values of the Chern numbers change by 1 when adjacent pairs of bands come into contact with or move apart from each other with increasing $M_{\mathrm{y}}$. A magnified view of Fig. \ref{gFig4}a for the $M_{\mathrm{y}}$ range of 0.139 meV to 0.142 meV is shown in Fig. \ref{gFig4}b to show that the Chern numbers of bands $iv$ and $v$ are both zero over a finite range of $M_{\mathrm{y}}$. The range spans from $M_{\mathrm{y}} = 0.1394$ meV, at which bands $iv$ and $v$ first touch at point $a$ on the $k-\phi_{\mathrm{U}}$ plane, to $M_{\mathrm{y}} = 0.1414$ meV when the two bands touch again at point $b$. At both of these points, a unit of Chern number is transferred between the two bands. 

As a different example, we can consider the transfer of Chern number between bands $vii$ and $viii$. Bands $vii$ and $viii$ initially overlap with each other. As $M_{\mathrm{y}}$ is increased, the two bands move apart (cf. Figs \ref{gFF5}a and \ref{gFF5}b). A band touching point occurs at point $d$ on the $k-\phi_{\mathrm{U}}$ plane (Fig. \ref{gFF3}e) at $M_{\mathrm{y}} = 0.1392$ meV, at which one unit of Chern number is transferred from band $vii$ to band $viii$. A second transfer of Chern number between bands $vii$ and $viii$ occurs at $M_y=0.23$ meV, at which a touching point appears at point $c$ on the $k-\phi_{\mathrm{U}}$ plane. 

One distinct characteristic that can be observed is the large value of Chern number that can be associated with a particular band, i.e., up to a value of 3. This is attributed to the presence of multiple bands (in principle, infinite in number) in the periodically curved nanowire system due to the folding of the Brillouin zone. The multiplicity of bands allows an individual band to successively accumulate units of Chern numbers as touching points occur between adjacent pairs of bands with the increase in $M_{\mathrm{y}}$. For example, the Chern number of band $viii$ in Fig. \ref{gFig4}a has an initial Chern number of 1 which increases in two successive steps to 3 with the increase in $M_{\mathrm{y}}$. Such a large Chern number states is not observed in typical two- or four-band topological systems such as the BHZ Hamiltonian for 2D topological insulators \cite{Science314_1757} due to their fewer number of bands.
The total number of charges pumped across any point in the infinite nanowire system by the adiabatic variation of $\phi_{\mathrm{U}}$ from 0 to $2\pi$ is given by the sum of the Chern numbers of the occupied bands. We therefore show the cumulative sum of the Chern numbers from the lowest energy band up to the highest occupied band index for each value of $M_{\mathrm{y}}$ in Fig. \ref{gFig4}c. These results show that for a fixed number of occupied bands, the number of charges pumped across the system can be varied simply by tuning the strength of the magnetization coupling. For example, when seven bands are occupied, the Chern number can be successively reduced from 3 to 2 and then to 1 by increasing  $M_{\mathrm{y}}$ from 0 to beyond 0.23 meV. Since the Chern number is linked to the number of pumped charges per adiabatic cycle of the sliding potential, one can in principle achieve a discrete and topologically robust charge transfer, by modulating the magnetization coupling. This may find use in a magnetic memory application in which in which the value of the bit is represented by the Chern number \cite{JAP117_17C739, APE4_094201, SciRep4_5123}.

 It is more practicable however to control the Fermi level of the system rather than the scenario of having a certain number of fully-occupied bands and having no partially occupied ones. Hence, in Fig. \ref{gFig4}d, the cumulative Chern number is plotted as a function of Fermi energy $E_{\mathrm{f}}$, taking into consideration all occupied states lying below $E_{\mathrm{f}}$ and including the contribution to the Chern number from the partially occupied bands. Here, the sums of the Chern numbers are not necessarily whole integers as the Fermi energy may cut across partially occupied bands. We observe that the boundaries between regions with different cumulative Chern numbers form a linear variation with respect to $M_{\mathrm{y}}$. This is due to the fact that these boundaries are dependent on the energy of the band minima (at which successive bands emerge and start to contribute to the Chern number), and these vary linearly with $M_{\mathrm{y}}$. Furthermore, the distinct regions in Fig. \ref{gFig4}d with different cumulative Chern number can be mapped to the corresponding region in Fig. \ref{gFig4}c. For example, the region around $E_\mathrm{f} = 4\ $ meV in Fig. \ref{gFig4}d indicated by the red dotted quadrilateral with the Chern number of 4 corresponds directly to the occupany of eight bands in Fig. \ref{gFig4}c. 

Besides modifying the strength of the magnetization coupling, an alternative and perhaps easier method to tune the Chern number is by varying its direction. We plot in Fig. \ref{gFig4}e and Fig. \ref{gFig4}f the cumulative Chern number as a function of the number of occupied bands and $E_f$ for the same system, except that the direction of the magnetization coupling is varied on the $x$-$y$ plane while keeping the coupling strength constant (at 0.3 meV), i.e., $\hat{M}=\cos(\phi_{\mathrm{M}})\hat{x} + \sin(\phi_{\mathrm{M}})\hat{y}$. As a result, the cumulative sum of the Chern numbers will also vary with the magnetization orientation denoted by $\phi_{\mathrm{M}}$. Fig. \ref{gFig4}f shows that when the Fermi energy is set to around 4 meV, the Chern number and hence the number of charges pumped per adiabatic cycle can be varied drastically between zero to four by varying the magnetization angle. Hence, by setting the $E_\mathrm{f}$ at an appropriate value and using $\phi_{\mathrm{M}}$ as an input variable, one can thus control the discrete number of pumped charges per phase period as a topologically robust output.
\section{Conclusions} 
In this work, we showed that the topological properties of a one-dimensional nanowire may be modified through the introduction of a periodic curvature, sliding potential and magnetization coupling. We described a scheme for obtaining the finite difference approximation to the Hamiltonian of a one-dimensional curved nanostructure in the presence of spin-orbit interaction (SOI). We showed that the initial dispersion with the SOI-induced split may be further modified with  the addition of the magnetization coupling. This has the effect of modifying the ordering and energy gap of the different bands. Further tuning of the dispersion is possible by introducing a sliding potential and varying the phase difference $\phi_{\mathrm{U}}$ between it and that of the periodic curvature. More importantly, the presence of $\phi_{\mathrm{U}}$ provides an additional parameter to obtain a well-defined invariant or Chern number which characterize the topological state of the system. By aligning the Chern number to the sliding potential, we are also able to link this invariant to a physical process, i.e., the number of discrete charges being pumped across the curved nanowire per adiabatic period of the sliding potential. The number of pumped charges therefore serve as a measurable indicator of the underlying topology of the system. We find that the multiplicity of bands in the system due to its periodicity and successive splits via SOI and magnetization coupling, allows the realization of states or bands with high Chern number. We showed that these distinct bands are connected by robust edge states which are localized at the boundaries, and which have the effect of transporting discrete states from one band to another, leading to shifts in the Chern number. The band and edge state profiles can be modulated by the strength and direction of the magnetization coupling, as well as the number of occupied bands (which in turn depends on the Fermi level) based on realistic parameter values. This opens the way for the system to be utilized for a multi-state memory application with topologically robust, discrete-valued readout value and having multiple levers of control.
\section{Acknowledgements}
We thank Prof. C.-R. Chang and R.-A. Chang at the National Taiwan University for helpful discussions during the course of this project. 
This work is supported by the Singapore National Research Foundation (NRF), Prime Minister’s Office, under its Competitive Research Programme (NRF CRP12-2013-01, NUS Grant No. R-263-000-B30-281), Ministry of Education (MOE) Tier-II Grant MOE2018-T2-2-117 (NUS Grant Nos. R-263-000-E45-112/R-398-000-092-112), MOE Tier-I FRC Grant (NUS Grant No. R-263-000-D66-114), and other MOE grants (NUS Grant Nos. C-261-000-207-532, and C-261-000-777-532).

\end{document}